\documentclass[prb, twocolumn,showpacs,superscriptaddress]{revtex4}
\usepackage{graphicx}
\usepackage{dcolumn}
\usepackage{bm}
\usepackage{amsmath,amsfonts}
\usepackage{latexsym}
\begin{document}
\title{Phase-dependent noise correlations in normal-superconducting structures}
\author{Markku P. V. Stenberg}
\email{markku.stenberg@tkk.fi}
\affiliation{%
Laboratory of Physics, Helsinki University of Technology, P.O. Box 4100,
FIN-02015 HUT, Finland}
\author{Pauli Virtanen}
\affiliation{Low Temperature Laboratory, Helsinki University of Technology,
P.O. Box 3500, FIN-02015 HUT, Finland}
\author{Tero T. Heikkil\"a}
\affiliation{Low Temperature Laboratory, Helsinki University of Technology,
P.O. Box 3500, FIN-02015 HUT, Finland}
\begin{abstract}
We study nonequilibrium noise correlations in diffusive normal-superconducting
structures in the presence of a supercurrent. We present a
parametrization for the quasiclassical Green's function in the first
order of the counting field $\chi$. This we employ to obtain the voltage 
and phase dependence of cross and autocorrelations and to describe the 
role played by the setup geometry. We find that the low-voltage behavior 
of the effective charge $q_{\rm eff}$ describing shot noise is a result of 
a competition between anticorrelation of Andreev pairs due to proximity 
effect and the depression of the local density of states. Furthermore, 
we show that the noise correlations are independent of the sign of the
supercurrent.6
\end{abstract}
\maketitle 
The charge transmitted through a disordered conductor in
unit time varies due to the quantum nature of the transport process.
This variation is characterized by the current distribution, whose
width at low temperatures is directly related to the shot
noise.\cite{blanter01} In
metallic conductors, the corrections induced by the quantum
coherence on the current and conductance distributions are 
small,\cite{muttalib03,stenberg06} but in a metal in contact to a
superconductor, more pronounced effects may be observed, e.g., in
the out-of-equilibrium noise 
experiments.\cite{jehl00,kozhevnikov00,lefloch03,reulet03} 
In the context of a
normal-superconducting (NS) two-terminal setup, the theoretical
endeavours have recently covered, e.g., the voltage dependence of
the shot noise \cite{belzig01d,stenberg02} and certain low-bias
anomalies.\cite{stenberg02,houzet04} Multiterminal structures have
been discussed in the incoherent 
regime,\cite{nazarov02,samuelsson02,belzig03a,virtanen06} and in the 
presence of 
a supercurrent, in a short junction,\cite{bezuglyi04} and for
specific values of a phase difference in a three-terminal setup. The
latter was described by a method based on a direct discretization of
the equations governing the full counting statistics [see
Eq.~(\ref{eq:usadel})].\cite{reulet03}

\begin{figure}
\includegraphics[width = 5.6cm]{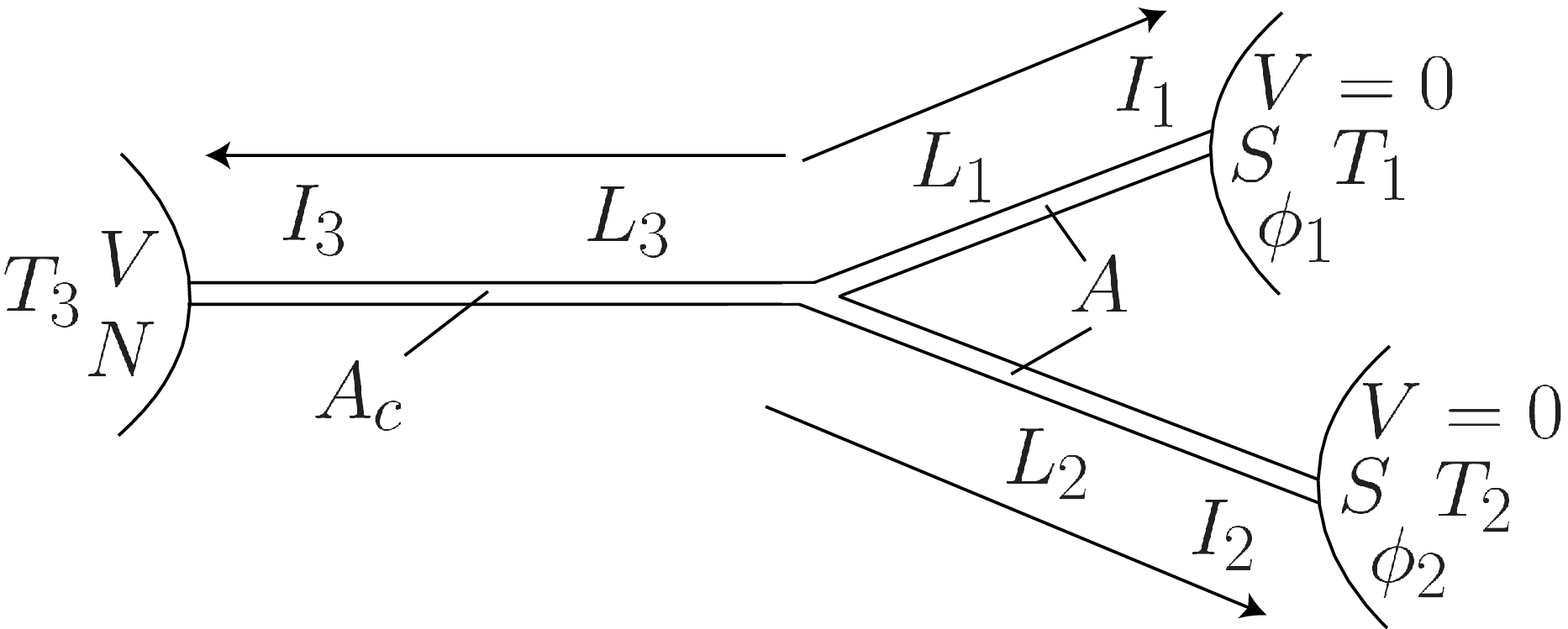}
\caption{Setup schematic studied in this work.}
\label{fig:cc_setup}
\end{figure}

In the presence of the superconducting proximity effect and at voltages of
the order of $E_T/e$, the effective charge $q_{\rm eff}$ characterizing the
magnitude of shot noise is lower than $2e$, the value corresponding to a
Cooper pair.\cite{belzig01d,stenberg02,reulet03} Here $E_T=\hbar D/L^2$ is 
the Thouless energy, with $D$
the diffusion constant and $L$ the wire length. Applying a supercurrent 
in a three-terminal structure (Fig.~\ref{fig:cc_setup}) allows one to study 
the nature of this proximity-induced change in $q_{\rm eff}$. 
For example, with this approach, we directly show that the lowering of 
$q_{\rm eff}$ is due to a 
competition of anticorrelation effects induced by the superconducting
proximity effect \cite{reulet03} and the depression of the local density of 
states.\cite{zhou98} In principle, the effect of an extra 
current in the system would be twofold: to tune the coherent effects and to 
induce its own correlations.  
Here we show that within our noninteracting 
model supercurrent does only the previous as all the correlations in the 
system are independent of the supercurrent sign.

The full counting statistics \cite{levitov96,nazarov99fcs,belzig03b}
has recently become the method of choice to calculate shot noise in
diffusive mesoscopic conductors but, to our knowledge, the cross
correlations in the presence of supercurrent have not been studied.
The statistics is accessed by performing a counting rotation
of the Green's function $\check{g}_0$ in one
of the terminals, using the counting field $\chi$.\cite{belzig03b}
To access the noise correlations, one has to expand the resulting
function $\check{g}(\chi) = \check{g}_0 -i(\chi/2)
\check{g}_1+\mathcal{O}(\chi^2)$ in powers of the counting field and
solve the problem in the first order in $\chi$. Within
quasiclassical formalism, the spectral quantities related to
$\check{g}_0$ may be represented by two parameters, $\theta$ and
$\phi$ characterizing the magnitude and phase of the pair
correlations. The distributions of the electrons and holes are
treated by dividing the functions into even and odd components with
respect to the Fermi surface, $f_T$ and $f_L$.\cite{belzig99} In
the absence of supercurrent, a parametrization for $\check{g}_1$ has
been given in Ref.~\onlinecite{houzet04}.
Here we present a parametrization for $\check{g}_1$ applicable also for
a finite supercurrent. Since the resulting essentially linear method is
considerably faster
than the previous ``direct discretization'' approach, we have been able to
extensively study the roles played by the setup geometry,
dissipative currents, coherence and phase gradients.

In our setup (Fig.~\ref{fig:cc_setup}), the two
superconducting terminals $T_1$ and $T_2$ are connected by diffusive
wires $1$ and $2$ and, at the central node, by a diffusive control
wire $3$ to a normal reservoir, $T_3$. The phase
difference $\Delta\phi=\phi_2-\phi_1$ 
between the superconductors may be generated by fabricating a
superconducting loop and applying a magnetic flux 
through the loop or by an external driving of supercurrent. The
lengths of the wires are denoted by $L_{1,2,3}$ and the currents
into the terminals $I_{1,2,3}$. We designate the cross section of
the control wire by $A_c$ and suppose that the other wires are
equally wide, with cross section $A$. The electric potential is
assumed to vanish in the superconductors. We assume good contacts at
the interfaces and a vanishing temperature and consider voltages
below the superconducting energy gap $V\ll \Delta/e$.

In the quasiclassical diffusive limit,
the triangular matrix $\check{g}_0=\check{g}(\chi=0)$ in
Nambu($\hat{\ }$)-Keldysh($\bar{\ }$)
space may be expressed through $\hat{R}$, $\hat{A}$,
and $\hat{K}$,
which may be parametrized using
$\hat{A} = -\hat{\tau}_3\hat{R}^{\dag}\hat{\tau}_3$,
$\hat{K} = \hat{R}\hat{h}-\hat{h}\hat{A}$, $\hat{h}=f_L+f_T\hat{\tau}_3$,
$\hat{R} = \cosh(\theta)\hat{\tau}_3+\sinh(\theta)
(\cos(\phi)i\hat{\tau}_2+\sin(\phi)i\hat{\tau}_1)$.\cite{belzig99}
Here $\theta$ characterizes the strength of the superconducting proximity
effect, $\phi$ is the superconducting phase, and $\hat{\tau}_i$ are the
Pauli matrices in Nambu space.
The counting field $\chi$, which we introduce in the normal
reservoir, appears as a gauge
transformation of the Green's function in the same terminal.\cite{belzig03b}
At a vanishing temperature, it suffices to concentrate only
on the energy regime $0<\varepsilon<eV$. In this case, without counting
rotation, one has in the normal terminal
$\check{g}_{N,0}=\check{g}_{N}(\chi=0)=
\hat{\tau}_3\otimes\bar{\sigma}_3+\hat{\tau}_0
\otimes(\bar{\sigma}_1+i\bar{\sigma}_2)$, where
$\bar{\sigma}_i$ are Pauli matrices in Keldysh space.
In the superconducting terminals we have
$\check{g}_{S}=(\cos(\phi)\hat{\tau}_2 + \sin(\phi)\hat{\tau}_1)
\otimes\bar{\sigma}_0$.
The generalized Green's functions
$\check{g}(\chi)$ obey the Usadel equation
\cite{usadel70}
similar to that in the zeroth order in $\chi$
\begin{eqnarray}
-\frac{\hbar D}{L G_D}\partial_x \check{J}(x) =
-i\varepsilon\left[\check{\tau}_3,\check{g}(x) \right],
\label{eq:usadel}
\end{eqnarray}
with $\check{J}(x) = -L G_D \check{g}(x)\partial_x \check{g}(x)$,
$\check{\tau}_3=\hat{\tau}_3\otimes \bar{\sigma}_0$.
Here $G_D$ is the normal-state conductance of the wire with length $L$,
$x$ is the coordinate along the wire, and $\varepsilon$ is the energy.
The Green's function satisfies the normalization condition
$\check{g}^2(\chi) = \hat{\tau}_0\otimes\bar{\sigma}_0$.
At the NS interfaces the Nazarov boundary conditions
\cite{nazarov99bc} for $\check{g}(\chi)$ hold.

We obtain the noise correlations from
\begin{equation}
S_{ij} \equiv \int_{-\infty}^{\infty}\mathrm{d}t
\langle \{\delta I_{i}(t),\delta I_{j}(0)\} \rangle =
-2ie \frac{\partial J_{i}(\chi)}{\partial\chi_{j}}|_{\chi = 0}.
\label{eq:noisedefin}
\end{equation}
Here we have $\qquad J(\chi)= -1/(8e)\int d\varepsilon \mathrm{Tr}
[\check{\tau}_K \check{J}(x)]$ and
$\delta I_{i}=I_{i}-\bar{I}_{i}$ is the deviation
of the current from its quantum mechanical expectation value.
In this article we take $j=3$ and thus with $i=3$ we get the noise
$S_{33}\equiv S$
and
with $i=1,2$ the cross correlations.
The effect of the current-voltage characteristics on the current
fluctuations may be eliminated by considering the effective
charge $q_{\rm eff}\equiv (3/2)|\textrm{d}S/\textrm{d}\bar{I}_{3}|$, where the
factor $3$ arises from the diffusive nature of the transport.
The effective
charge yields information on the charge transferred and also on the
energy-dependent correlations between charge transfers in the transport
process. The matrix current in the first order in $\chi$
\begin{equation}
\check{J}^{(1)}(x) \equiv  -2i\partial_{\chi}\check{J}(x)_{|\chi=0} 
= -L G_D(\check{g}_0\partial_x \check{g}_1
+\check{g}_1\partial_x \check{g}_0)
\label{eq:firstordercurrentdef}
\end{equation}
is defined so that the Usadel equation in the first order in $\chi$
is identical to Eq.~(\ref{eq:usadel})
with the substitution $\check{g}\rightarrow \check{g}_1,
\check{J}\rightarrow \check{J}^{(1)}$.
The Nazarov boundary conditions for $\check{J}^{(1)}$ are given by
\cite{houzet04,misprint04}
\begin{eqnarray}
&\check{J}^{(1)} =
- 2 G_B \frac{\sum_n T_n\check{A}\check{B}\check{A}}{\sum_n{T_n}},
\ \check{A}=[4+T_n(\{\check{g}_0,\check{g}_S\}-2)]^{-1}, 
\nonumber\\&
\check{B}=4[\check{g}_1,\check{g}_S]
+ 2T_n (\check{g}_S \check{g}_0 \check{g}_1 \check{g}_S -
\check{g}_0 \check{g}_1- [\check{g}_1,\check{g}_S]).
\label{eq:firstorderbc}
\end{eqnarray}
Here $\{ T_n\}$ are the eigenvalues of the transmission matrix through the
interface, with conductance $G_B = e^2\sum_{n}T_n/(\pi\hbar)$. Below, we
assume a transparent contact, $G_B \gg G_D$.
The normalization of $\check{g}(\chi)$ implies
$\{\check{g}_0(x),\check{g}_1(x)\}=0$. This is readily satisfied by
introducing the change of the variables
$\check{g}_1(x)=[\check{g}_0(x),\check{\phi}(x)]$.

We find a parametrization for $\check{\phi}(x)$ valid also in the presence
of a supercurrent:
\begin{equation}
\check{\phi}=\left(
\begin{array}{c c}
\hat{r} & \hat{k}\\
\hat{l} & \hat{a}
\end{array}\right)
=\left(
\begin{array}{c c}
r_{1}\hat{\tau}_1+r_{3}\hat{\tau}_3 & k_0\hat{\tau}_0+k_3\hat{\tau}_3\\
f_L\hat{\tau}_0-f_T\hat{\tau}_3 & r_{1}^*\hat{\tau}_1-r_{3}^*\hat{\tau}_3
\end{array}\right),
\label{eq:para2}
\end{equation}
with $r_1=r_{11}+r_{12}i,\ r_3=r_{31}+r_{32}i$, and $r_{11}$,
$r_{12}$, $r_{31}$, $r_{32}$, $k_{0}$, $k_{3}$ $\in \mathbb{R}$. With this
parametrization, $\chi$ has to be generated in the normal terminal
and an arbitrary number of superconducting terminals be at zero
potential. Because of the specific matrix structure of the Usadel
equation, $\hat{l}$, $\hat{r}$, $\hat{a}$, and $\hat{k}$ may be
solved consecutively, and $\hat{r}$ and $\hat{a}$ are related by the
retarded-advanced symmetry. At the NS interface,
Eq.~(\ref{eq:firstorderbc}) yields the boundary conditions 
\begin{eqnarray}
&r_1 = r_3 = k_3 = 0,\\ 
&k_0' = \frac{2\sin\phi_1(f_T' r_{12}'\theta_{1}'+2 f_T' r_{11}'\theta_{2}'+
\phi_{2}' r_{12}'\theta_{1}'f_L-\phi_{1}'r_{12}'\theta_{2}'f_L)}{
\phi_2'^2-\theta_2'^2}
\end{eqnarray} 
while at the normal-terminal interface the boundary conditions read
$\check{g}_1=[\check{\tau}_K,\check{g}_{N,0}]$, and one may choose,
e.g., $r_1 = r_3 = k_0 = 0$, $k_3 = -1$. 

We obtain two differential equation systems which can be solved 
consequtively, one for the retarded part (upper left $2\times 2$ matrix) and 
one for the Keldysh part (upper right $2\times 2$ matrix) of 
Eq.~(\ref{eq:usadel}) in the first order in $\chi$. 
Not all the coefficients in the retarded part are 
independent but the equations take the form
\begin{eqnarray}
&&\mathcal{R}_{11}^{(2)}r_{11}''+\mathcal{R}_{11}^{(1)}r_{11}'
+\mathcal{R}_{11}^{(0)}r_{11}+\mathcal{R}_{12}^{(2)}r_{12}''
+\mathcal{R}_{12}^{(1)}r_{12}'\nonumber\\
&&+\mathcal{R}_{12}^{(0)}r_{12}+\mathcal{R}_{31}^{(2)}r_{31}''
+\mathcal{R}_{31}^{(1)}r_{31}'+\mathcal{R}_{32}^{(2)}r_{32}''
+\mathcal{R}_{32}^{(1)}r_{32}'=\mathcal{C}_{1},
\nonumber\\
&&-\mathcal{R}_{12}^{(2)}r_{11}''-\mathcal{R}_{12}^{(1)}r_{11}'-
\mathcal{R}_{11}^{(0)}r_{11}+\mathcal{R}_{11}^{(2)}r_{12}''
+\mathcal{R}_{11}^{(1)}r_{12}'\nonumber\\
&&+\mathcal{R}_{11}^{(0)}r_{12}-\mathcal{R}_{32}^{(2)}r_{31}''
-\mathcal{R}_{32}^{(1)}r_{31}'+\mathcal{R}_{31}^{(2)}r_{32}''
+\mathcal{R}_{31}^{(1)}r_{32}'=\mathcal{C}_{2}, 
\nonumber\\
&&\mathcal{R}_{31}^{(2)}r_{11}''+\mathcal{P}_{11}^{(1)}r_{11}'
+\mathcal{P}_{11}^{(0)}r_{11}+\mathcal{R}_{32}^{(2)}r_{12}''
+\mathcal{P}_{12}^{(1)}r_{12}'\nonumber\\
&&+\mathcal{P}_{12}^{(0)}r_{12}+\mathcal{P}_{31}^{(2)}r_{31}''
+\mathcal{P}_{31}^{(1)}r_{31}'+\mathcal{P}_{32}^{(2)}r_{32}''
+\mathcal{P}_{32}^{(1)}r_{32}'=\mathcal{C}_{3},
\nonumber\\
&&-\mathcal{R}_{32}^{(2)}r_{11}''-\mathcal{P}_{12}^{(1)}r_{11}'
-\mathcal{P}_{12}^{(0)}r_{11}+\mathcal{R}_{31}^{(2)}r_{12}''
+\mathcal{P}_{11}^{(1)}r_{12}'\nonumber\\
&&+\mathcal{P}_{11}^{(0)}r_{12}-\mathcal{P}_{32}^{(2)}r_{31}''
-\mathcal{P}_{32}^{(1)}r_{31}'+\mathcal{P}_{31}^{(2)}r_{32}''
+\mathcal{P}_{31}^{(1)}r_{32}'=\mathcal{C}_{4}.
\nonumber\\
\label{eq:retardedeqs}
\end{eqnarray}
Here $\mathcal{R}_{ij}^{(k)},\ \mathcal{P}_{ij}^{(k)}, \mathcal{C}_{i}\in 
\mathbb{R}$ 
depend on $\theta,\phi,f_{L,T}$ and their derivatives.
The first and second lines of Eq.~(\ref{eq:retardedeqs}) are obtained by 
operating with ${\rm Re}[{\rm Tr}(\hat{\tau}_1\cdot)]$ and 
${\rm Im}[{\rm Tr}(\hat{\tau}_1\cdot)]$ on the retarded part of
Eq.~(\ref{eq:usadel}), respectively, while 
${\rm Re}[{\rm Tr}(\hat{\tau}_3\cdot)]$ and 
${\rm Im}[{\rm Tr}(\hat{\tau}_3\cdot)]$
yield the equations on the third and fourth lines.
The full expressions, however, are too long to write here. 
The Keldysh part obeys two coupled 
differential equations
\begin{eqnarray}
\mathcal{K}_{0}^{(2)}k_{0}''+\mathcal{K}_{0}^{(1)}k_{0}'+
\mathcal{K}_{3}^{(2)}k_{3}''+\mathcal{K}_{3}^{(1)}k_{3}'=\mathcal{S}_{1},
\nonumber\\
-\mathcal{K}_{0}^{(2)}k_{0}''+\mathcal{Q}_{0}^{(1)}k_{0}'+
\mathcal{Q}_{3}^{(2)}k_{3}''+\mathcal{Q}_{3}^{(1)}k_{3}'=\mathcal{S}_{2}.
\label{eq:keldysheqs}
\end{eqnarray}
Here $\mathcal{K}_{0,3}^{(1,2)}$, $\mathcal{Q}_{0,3}^{(1,2)},$ 
$\mathcal{S}_{1,2}$  $\in \mathbb{R}$ depend on 
$\theta,\phi,f_{L,T},r_{1,3}$ and their derivatives.
The first and second line of Eq.~(\ref{eq:keldysheqs}) are obtained from the 
Keldysh part of Eq.~(\ref{eq:usadel}) in the first order in $\chi$ by taking 
the traces ${\rm Re}[{\rm Tr}(\cdot)]$ and 
${\rm Re}[{\rm Tr}(\hat{\tau}_3\cdot)]$, respectively.

Putting all together, the spectral equations for $\theta,\phi$ and the 
kinetic equations for $f_L,f_T$ are first solved, then
Eq.~(\ref{eq:retardedeqs}) for $r_1,r_3$, and thereafter 
Eq.~(\ref{eq:keldysheqs}) for $k_0,k_3$. Equations 
(\ref{eq:noisedefin}) and (\ref{eq:firstordercurrentdef}) yield a
lengthy expression for noise correlations in terms of the parameters 
$\theta, \phi, f_L, f_T, r_1, r_3, k_0, k_3$
into which the values of these parameters are finally substituted.
Essentially because of the finite "coherence" parameter $r_1$,
noise deviates from its incoherent value \cite{blanter01},
and due to a finite supercurrent, $r_3$ may be finite.
However, in up-down (1-2) symmetric structures, 
$r_3$ and $k_0$ always vanish in 
the control wire.
We have developed a computer code to solve numerically these equations and 
present the results below. We first consider up-down symmetric setups and
then the influence of breaking this symmetry.

\begin{figure}
\includegraphics[width = 8cm]{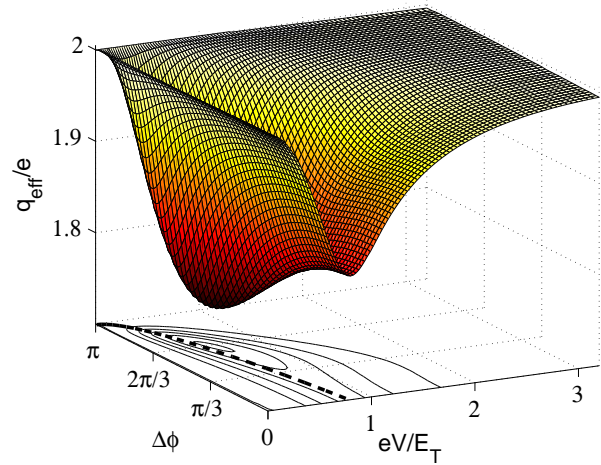}
\caption{(Color online): Effective charge $q_{\rm eff}$ vs voltage
and $\Delta\phi$. The voltage at which the minimum
$q_{\rm eff}$ is obtained follows the phase-dependent minigap in the 
SNS system (bold dashed black curve).
Note that the junction undergoes a
$\pi$-transition, in which the supercurrent changes its sign,
at $eV \approx 2 E_T$.\cite{heikkila02,huang02}}
\label{fig:qeff_vs_E_phi}
\end{figure}

The full phase and voltage dependence of $q_{\rm eff}$ with $L_{1,2,3}=L$
is illustrated in Fig.~\ref{fig:qeff_vs_E_phi}. 
The $I-V$ characteristics
in such a structure may be calculated from $\check{J}(\chi=0)$
in a way explained, e.g., in Ref.~\onlinecite{belzig99}. 
Our results for $\Delta
\phi=0,\pi$ coincide with those in Ref.~\onlinecite{reulet03} and we
obtain a minimum of $q_{\rm eff}$ at about $\Delta\phi = 0.63 \pi$
(corresponding to the maximum of the spectral supercurrent \cite{belzig99}),
$eV = 0.5\ E_{T}$. 
The nonmonotonic voltage dependence of $q_{\rm eff}$ may be understood
by studying the Andreev reflection eigenvalue density of a diffusive
wire, which at $V=0$ takes the form of the Dorokhov 
distribution.\cite{samuelsson04} 
The behavior of the noise parameters as a
function of $\theta$ suggests that the returning of $q_{\rm eff}$ to
$2e$ at low voltages may also be attributed to the depression of the
local density of states at low energies.\cite{zhou98} This is
illustrated by the fact that the voltage at which the minimum of
$q_{\rm eff}$ is obtained follows the phase-dependent minigap in the 
superconductor-normal metal-superconductor (SNS) system.
\begin{figure}
\includegraphics[width = 8cm]{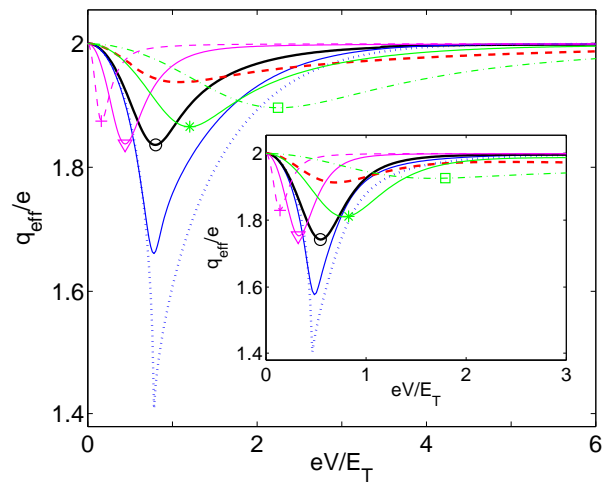}
\caption{(Color online): Effective charge $q_{\rm eff}$ vs $eV$ for
$\Delta \phi=0$ (main figure) and $\Delta \phi=2\pi/3$ (inset). 
The cross section $A_c$ is close to $0$ (blue dotted), $A/4$ (blue
solid), $A$ (black bold $\circ$), $4A$ (red dashed) and for these curves 
$L_{1,2,3} = L$. The values for $L_3$ are $1/8$
(green dash-dotted $\Box$), $1/2$ (green solid $\ast$), $2$ (magenta
solid $\bigtriangledown$), $4$ (magenta dashed $+$) times
$L_{1,2}=L$ with $A_c = A$; $E_T = \hbar D/L^2$. For $\Delta\phi=2\pi/3$,
the minima of $q_{\rm eff}$ occur at lower $V$ and
except for the curves for $A_c=0$ and $L_3 = L/8$, 
the dips are deeper than for $\Delta \phi=0$.
Note the different energy scales on the x-axes in the inset and in the main
figure.
} \label{fig:ac_dissipation}
\end{figure}
\begin{figure}
\includegraphics[width = 8cm]{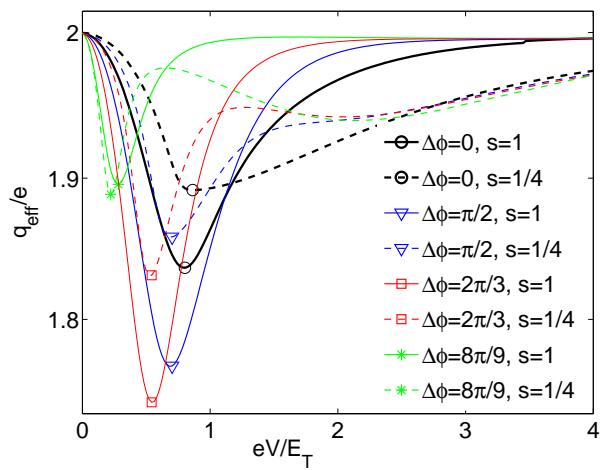}
\caption{(Color online): Effective charge $q_{\rm eff}$ for
different values of phase difference in left-right symmetric ($s=1$)
and asymmetric ($s=1/4$) structures.} \label{fig:ac_asymmetry}
\end{figure}
\begin{figure}
\includegraphics[width = 8cm]{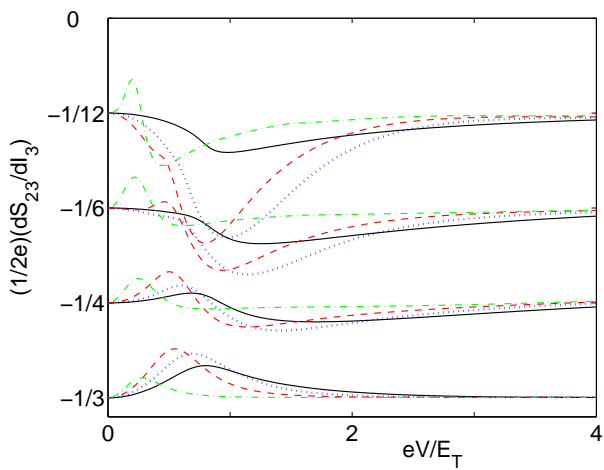}
\caption{(Color online): Normalized differential cross correlations
$(1/2e)(\mathrm{d}S_{23}/\mathrm{d}\bar{I}_3)$ vs $eV$, for $\Delta
\phi=0$ (black solid), $\Delta \phi=\pi/2$ (blue dotted), $\Delta \phi=2\pi/3$
(red dashed), $\Delta \phi=8\pi/9$ (green dash-dotted) and 
$s=1/4, 1/2, 3/4, 1$ for the curves from top to bottom. 
The cross correlations
related to the opposite wire are directly obtained from 
$S_{13}=-S_{23}-S_{33}$. 
The results for $S_{23}$ and the symmetry parameter $s$ are identical for 
those for $S_{13}$ and $2-s$ by the symmetry of $S_{ij}$ under the reversal 
of the phase gradient.} 
\label{fig:cc_asymmetry}
\end{figure}

In Fig.~\ref{fig:ac_dissipation}, $q_{\rm eff}$ vs $eV$ is plotted
for a vanishing $\Delta \phi$ and for different cross sections $A_c$
and lengths $L_3$ of the control wire, with given $E_T=\hbar D/L^2$.
The influence of a finite $\Delta\phi$ is exemplified in the inset,
where the results for the same parameters as in the 
Fig.~\ref{fig:ac_dissipation}, but for $\Delta \phi = 2\pi/3$, have
been plotted. 
The behavior of
the supercurrent in somewhat similar situations has been studied
earlier in Refs.~\onlinecite{heikkila02} and \onlinecite{huang02}. 
For $A_c \rightarrow 0$, the voltage at which the minimum $q_{\rm eff}$ is 
obtained exactly coincides with the phase-dependent minigap in the SNS system.
Generally, enlarging the width of the control wire or varying the lengths 
away from the symmetric case $L_{1,2,3}=L$ tends to make the
dip shallower. With $L_{1,2}=L$ and $L_3$ smaller
(larger) than $L$, the minimum in $q_{\rm eff}$ is shifted to higher
(lower) voltages. This is in agreement with the conclusion that the
dip is caused by the anticorrelation between subsequent Andreev
pairs as the correlations of the pairs manifest themselves
at the length scale $\sim (\hbar D/\varepsilon)^{1/2}$. 

Under the sign reversal of $\Delta\phi$, at energies
below $eV$, the dissipative and superconducting parts of the
spectral charge current $j_T$ remain invariant.\cite{heikkila03} A direct 
calculation shows that also $S_{ij}$ is invariant under the sign reversal of
$\Delta \phi$ (also in asymmetric structures). This means that the
quasiparticle current is in no way correlated with the supercurrent
flowing in the system, although the presence of the latter changes
the correlations in the previous. Hence in the left-right symmetric
structures, we have $S_{13}=S_{23}=-S_{33}/2$ through current
conservation. Note that if the supercurrent
would be replaced by a dissipative current, the cross correlations
would depend on the relative signs between $\bar{I}_3$ and the
"circulating" current $\bar{I}_1-\bar{I}_2$. 
 In order to study the effect of asymmetry, we
introduce a symmetry parameter, $0< s < 2$, measuring the distance
between the wire $3$ and $T_1$ such that
$L_{1}/s=L_{2}/(2-s)=L_3=L$. The symmetric system described above
corresponds thus to $s=1$. The same results for $q_{\rm eff}$ apply
for the symmetry parameters $s$ and $2-s$ as $q_{\rm eff}$ is
invariant under the change of the sign of $\Delta \phi$. Hence we
restrict ourselves to $s\le 1$ without loss of generality. In
Fig.~\ref{fig:ac_asymmetry}, we have calculated $q_{\rm eff}$, i.e.,
a normalized autocorrelation function by varying $s$ and $\Delta
\phi$. 
With decreasing $s$, the coherence in the control wire
increases as the other superconducting terminal is brought closer to
it. 
At voltages higher than $E_T/e$, but in the region where the
coherence is not fully suppressed, the enhanced coherence gives rise
to the long tails with $q_{\rm eff}< 2e$ in
Fig.~\ref{fig:ac_asymmetry}. 
However, at voltages near the minimum
of $q_{\rm eff}$, decreasing $s$ suppresses the dip. 

Figure~\ref{fig:cc_asymmetry} represents the normalized differential
cross correlations $(1/2e)(\textrm{d}S_{23}/\textrm{d}\bar{I}_{3})$. 
In the lowest curves for the symmetrical case,
$s=1$, we have $(1/2e)(\textrm{d}S_{23}/\textrm{d}\bar{I}_{3}) = -q_{\rm
eff}/6e$. In
the incoherent region $eV\gg E_T$, the absolute value of $S_{23}$
diminishes linearly with decreasing $s$, as one may anticipate on
the basis of the Kirchoff rules. 
In the coherent regime, 
$\textrm{d} S_{23}/\textrm{d}\bar{I}_3$ reflects, e.g., 
the relative changes in $\textrm{d}\bar{I}_2/\textrm{d}V$ and 
$\textrm{d}\bar{I}_3/\textrm{d}V$ with $V$. 
In the symmetric case, these 
relative changes have equal magnitudes.
If the former exhibits a larger change than the latter, 
$(1/2e)(\textrm{d}S_{23}/\textrm{d}\bar{I}_3)$ may obtain larger negative 
values than in the incoherent 
regime. 
With a finite supercurrent these dips correspond to the processes in 
which two electrons are injected from the normal reservoir and an 
Andreev pair enters the superconductor $T_2$.

In conclusion, we have found a physically transparent and
computationally efficient way to calculate the full phase and
voltage dependence of the noise correlations in mesoscopic diffusive
wires in the presence of supercurrent. 
We found that the strength of the anticorrelations between the
Andreev pairs flowing in the structure is closely related to the magnitude 
(but not to the sign) of the spectral supercurrent and the variations in the 
local density of states.

We acknowledge W. Belzig, M. Houzet, and F. Pistolesi for
discussions and Center for Scientific Computing for computing resources.
M.P.V.S. acknowledges the financial support of Magnus Ehrnrooth Foundation
and the Foundation of Technology (TES, Finland). P. V. thanks the Finnish 
Cultural Foundation and T.T.H. the Academy of Finland for funding.


\begin{thebibliography}{27}
\bibitem[{There is a misprint in this formula in
  Ref.~\onlinecite{houzet04}()}]{misprint04}
There is a misprint in this formula in Ref.~\onlinecite{houzet04}.
\end{thebibliography}
\end{document}